\documentclass[12pt]{article}
\usepackage{amssymb,amsmath,amscd,amsthm}
\usepackage[utf8x]{inputenc}
\usepackage[T2A]{fontenc}

\usepackage{graphicx,psfrag}

\newtheorem{theorem}{Theorem}[section]

\newtheorem{remark}{Remark}[section]
\newtheorem{lemma}{Lemma}[section]
\newtheorem{proposition}{Proposition}[section]

\newtheorem{problem}{Problem}[section]

\font\bbc=msbm10 scaled 1200

\numberwithin{equation}{section}

\newcommand{\R}{\mbox {\bbc R}}

\newcommand{\sgn}{\text{sgn}}
\newcommand{\const}{\text{const}}
\newcommand{\Const}{\text{Const}}

\date{}

\begin{document}
 \author{ Evgeny L. Lakshtanov,\thanks{Department of Mathematics, Aveiro University, Aveiro 3810, Portugal.  This work was supported by Portuguese funds through the CIDMA - Center for Research and Development in Mathematics and Applications, and the Portuguese Foundation for Science and Technology (``FCT--Fund\c{c}\~{a}o para a Ci\^{e}ncia e a Tecnologia''), within project PEst-OE/MAT/UI4106/2014 (lakshtanov@ua.pt)} \and Roman G. Novikov\thanks{CNRS (UMR 7641), Centre de Math\'{e}matiques
Appliqu\'{e}es, \'{E}cole Polytechnique, 91128, Palaiseau, France, and IEPT RAS, 117997 Moscow, Russia (novikov@cmap.polytechnique.fr)},
 \and Boris R. Vainberg\thanks{Department
of Mathematics and Statistics, University of North Carolina,
Charlotte, NC 28223, USA. The work was partially supported  by the NSF grant DMS-1410547 (brvainbe@uncc.edu).}}

\title{A global Riemann-Hilbert problem for two-dimensional inverse scattering at fixed energy}

\maketitle

\begin{abstract}
We develop the Riemann-Hilbert problem approach to inverse scattering for the two-dimensional Schrodinger equation at fixed energy.
We obtain global or generic versions of the key results of this approach for the case of positive energy and compactly supported potentials.
In particular, we do not assume that the potential is small or that Faddeev scattering solutions do not have singularities (i.e. we allow the Faddeev exceptional points to exist). Applications of these results to the Novikov-Veselov equation are also considered.
\end{abstract}


\section{Introduction}
We consider the two-dimensional Schrodinger equation
\begin{equation}\label{fir}
(-\Delta+v)\psi(x ) =E \psi( x ), ~  x  \in \R^2, \quad E>0,
\end{equation}
where
\begin{equation}\label{fir2}
\mbox {$v$ is a real-valued sufficiently regular function on $\mathbb R^2$ with sufficient decay at infinity. }
\end{equation}
Actually, in the present work the assumptions (1.2) are specified in the sense that $v$ is a real-valued, bounded, compactly supported function on $\mathbb R^2$.

For equation (\ref{fir}) we consider the classical scattering solutions $\psi^+(x,k), ~k \in \mathbb R^2, ~k^2=E,$ specified by the following assymptotics
\begin{equation}\label{fir3}
\psi^+(x,k)=e^{ikx} + i \pi\sqrt{2\pi} e^{-i\pi/4}  \frac{e^{i|k||x|}}{\sqrt{|k||x|}} f\left (k,|k|\frac{x}{|x|} \right ) + o \left (\frac{1}{\sqrt{|x|}} \right), \quad |x| \rightarrow \infty,
\end{equation}
for some a priori unknown $f$. Function $f=f(k,l)$ on
\begin{equation}\label{fir4}
\mathcal M_E = \{k,l \in \mathbb R^2 ~ : ~ k^2=l^2=E \}
\end{equation}
arising in (\ref{fir3}) is the classical scattering amplitude for equation (\ref{fir}).

In order to determine $\psi^+$ and $f$ from $v$ one can use the Lipmann-Schwinger integral equation (\ref{eq1}) and the integral formula (\ref{eq1303}) in section 2; see e.g. \cite{RN2}.

In this work we continue, in particular, studies on the following inverse scattering problem for equation (\ref{fir}) under assumptions (\ref{fir2}):

\begin{problem}\label{prob11} Given scattering amplitude $f$ on $\mathcal M_E$ at fixed $E>0$, find the potential $v$ on $\mathbb R^2$.
\end{problem}

When $v$ is compactly supported, that is
\begin{equation}\label{fir5}
\mathrm{supp} ~  v \subset  D,
\end{equation}
where $D$ is an open bounded domain in $\mathbb R^2$, we consider also the Dirichlet-to-Neumann map $\Phi(E)$ for equation (\ref{fir}) in $D$. We recall that this map is defined via the relation
\begin{equation}\label{fir6}
\left . \frac{ \partial } {\partial \nu} \psi \right |_{\partial D} = \Phi(E) \left (\psi|_{\partial D} \right )
\end{equation}
fulfilled for all sufficiently regular solutions $\psi$ of (\ref{fir}) in $D \cup \partial D$, where $\nu$ is the external normal vector to $\partial D$.
Considering $\Phi(E)$, we assume also that
\begin{equation}\label{fir22}
\mbox{$E$ is not a Dirichlet eigenvalue for the operator $-\Delta+v$ in $D$.}
\end{equation}

It is well known (see \cite{novikov88}) that, under assumptions (\ref{fir2}), (\ref{fir5}), problem \ref{prob11} is closely related  with the following inverse boundary value problem for equation (\ref{fir}) in $D$:

\begin{problem}\label{prob12}  Given $\Phi(E)$ at fixed $E>0$, find  $v$.
\end{problem}

Problems \ref{prob11}, \ref{prob12} have a long history and there are many important results on these problems; see \cite{eskin}, \cite{g2000}, \cite{RN2}, \cite{novikov2015}, \cite{nov1} and references therein in connection with problem \ref{prob11} and \cite{b}, \cite{novikov88}, \cite{nov1}, \cite{sant} in connection with problem \ref{prob12}.

The approach of the present work to problems \ref{prob11}, \ref{prob12} is based, in particular,  on properties of the Faddeev  exponentially increasing solutions for  equation (\ref{fir}). We recall that the Faddeev solutions $\psi(x,k), k \in \mathbb C^2 \backslash \mathbb R^2, k^2=E$, of equation (\ref{fir}) are specified by
\begin{equation}\label{fir7}
\psi(x,k) = e^{ikx}(1+o(1)), ~ |x| \rightarrow \infty;
\end{equation}
see e.g. \cite{novikov88}.

In order to determine $\psi$ from $v$ one can use  the Lipmann-Schwinger-Faddeev integral equation (\ref{19JanA}) in section \ref{sec2}.

In the present work, under assumptions (\ref{fir2}), (\ref{fir5}), we reduce problems \ref{fir},\ref{fir2} to some global generalized Riemann-Hilbert-Manakov problem for the classical scattering solutions $\psi^+$ and the Faddeev solutions $\psi$ for equation (\ref{fir}); see problem \ref{prob32} in section \ref{sec3}.  A prototype of this global Riemann-Hilbert-Manakov problem for the case of equation (\ref{fir}) with $E<0$ was considered in section 8 of \cite{RN2}.

The term "global" means, in particular, that the kernels of our Riemann-Hilbert-Manakov problem have no singularities, even if there are the Faddeev exceptional points at fixed $E$.   After that we reduce our Riemann-Hilbert problem to a Fredholm linear integral equation of the second type; see theorem \ref{thmir1} and proposition \ref{prop41} in section \ref{sec4}.

As a result we obtain, in particular, a new generic reconstruction method for problems \ref{fir},~\ref{fir2}; see proposition \ref{nchg1} and remarks \ref{nchg2}, \ref{nchg3} in section \ref{sec4}.

In particular, our reconstruction from the Faddeev generalized
scattering data is reduced to formulas (\ref{smir5}), (\ref{smir7}), (\ref{mir3}), (\ref{mir4}), (\ref{mir9}), (\ref{frl1}), (\ref{frl1B}), integral
equations (\ref{mir6})-(\ref{mir8}),
(\ref{frl2}),(\ref{fr12B}) and formulas (\ref{aprilA1}),(\ref{aprilA2}),(\ref{aprilA3}),(\ref{mir11}),


Note that  the approach of the present work goes back to the soliton theory, see \cite{ayf}, \cite{fa83}, \cite{gn85}, \cite{gn86}, \cite{manakov81}.
The first applications of this approach to problems (\ref{fir}), (\ref{fir2}) were given in \cite{gm}, \cite{novikov86}, \cite{novikov88}, \cite{RN2}.
Actually, the main result of the present work consists in a globalization of this approach to problems (\ref{fir}), (\ref{fir2}).

The reconstruction method of the present work uses properly generalized scattering data for small and large values of the complex spectral parameter at fixed energy and, therefore, is considerably more stable, generically, than the reconstruction method of \cite{b} based exclusively on properties of some generalized scattering data for large values of complex spectral parameter. Generically, stability estimates of \cite{novikov10} obtained using ideas of \cite{al}, \cite{b} can be improved using results of the present work to estimates like in \cite{sant}, but without the assumptions that some norm of potential $v$ is sufficiently small in comparison with fixed $E$. This issue will be presented in detail elsewhere.

In addition, in contrasts with  \cite{b}, results of the present work admit application to solving the Cauchy problem for the Novikov-Veselov equation (\cite{Manakov76}, \cite{veselovnovikov})
\begin{eqnarray}\label{zir5}
&\partial_t v =4\Re(4\partial_z^3v+\partial_z(vw)-E\partial_zw), \\ \nonumber
&\partial_{\bar z}w=-3\partial_zv,\ \ v=\bar v,\ \ E >0,\\ \nonumber
& v=v(x,t),\ \ w=w(x,t),\ \ x=(x_1,x_2)\in\R^2,\ \ t\in\R,
\end{eqnarray}
with compactly supported  $v(x,t=0)$. Here, we used the following notations:
\begin{eqnarray}\label{hir1}
\partial_t={\partial \over \partial t}, \quad
\partial_z={1\over 2}\left({\partial\over \partial x_1}-i{\partial\over \partial x_2}\right),\quad
\partial_{\bar z}={1\over 2}\left({\partial\over \partial x_1}+i{\partial\over \partial x_2}\right).
\end{eqnarray}
These applications are indicated in section \ref{sec5} of the present work and will be presented in detail elsewhere.

\section{Preliminary results of direct scattering}\label{sec2}
\subsection{Classical scattering functions}\label{sec21}
We recall that for the classical scattering functions $\psi^+$ and $f$ for equation (\ref{fir}) the following Lipmann-Schwinger integral equation (\ref{eq1}) and the integral formula (\ref{eq1303}) hold:

\begin{equation}\label{eq1}
\psi^+(x,k) =e^{ikx } +
\int_{y \in \mathbb R^2}  G^+(x-y,\sqrt{E}) v(y) \psi^+(y,k)dy,
\end{equation}
$$
G^+(x,\sqrt{E}) =- \frac{1}{(2\pi)^2}  \int_{\mathbb R^2}\frac{e^{i\xi x}d\xi}{|\xi|^2 -E -i0}  = -\frac{i}{4}H^1_0(|x| \sqrt{E}),
$$
\begin{equation}\label{eq1303}
f(k,l) = \frac{1}{(2\pi)^2}  \int_{\mathbb R^2} e^{-ily} v(y) \psi^+(y,k)dy,
\end{equation}
where $x,k,l \in \mathbb R^2$, $k^2=l^2=E>0$, $H^1_0$ is the Hankel function of the first type; see e.g. \cite{RN2}. In addition, it is known that  equation (\ref{eq1}) is uniquely solvable with respect to $\psi^+(\cdot,k) \in L^\infty(\mathbb R^2)$ at fixed $k$, under conditions (\ref{fir2}) and, in particular, under the conditions that
\begin{equation}\label{fir9}
\mbox{$v=\overline{v} \in L^\infty(\mathbb R^2)$, \quad $\mathrm{supp }~ v \subset D,$ }
\end{equation}
where $D$ is an open bounded domain in $\mathbb R^2$; see e.g. \cite{berezinshubin} for a proof of a similar result in three dimensions.

Let
\begin{equation}\label{fir10}
\mathbb S^1_r = \{ \zeta \in \mathbb R^2 ~ : ~ \zeta^2 = r^2\}, ~ r>0,
\end{equation}
\begin{equation}\label{fir11}
\Sigma_E = \{ \zeta \in \mathbb C^2 ~ : ~ \zeta^2 = E\}, ~ E>0,
\end{equation}
\begin{equation}\label{fir14}
\Sigma_{E,\rho} = \{ \zeta \in \Sigma_E  ~ : ~ | \Im \zeta | \geq \rho\}, ~ E>0, ~ \rho >0,
\end{equation}
and let
\begin{equation}\label{fir14B}
\chi_{E,\rho} \mbox{ be the characteristic function of } \Sigma_{E,\rho} \mbox{ in } \Sigma_{E}.
\end{equation}

Note that $\mathcal M_E = \mathbb S^1_{\sqrt{E}}\times \mathbb S^1_{\sqrt{E}},$ where $\mathcal M_E$ is defined by (\ref{fir4}).

It is well known that, under conditions (\ref{fir2}), (\ref{fir5}),
\begin{equation}\label{fir8}
\mbox{$\psi^+(x,k)$ admits a holomorphic extension in $k$ from $\mathbb S^1_{\sqrt{E}}$ to $\Sigma_E$ at fixed $x$}
\end{equation}
and
\begin{equation}\label{fir9B}
\mbox{$f(k,l)$ admits a holomorphic extension in $(k,l)$ from $\mathcal M_E$ to $\Sigma_E \times \Sigma_E$ }
\end{equation}
with possible exponential increasing at infinity in complex domain.

As a corollary, $f$ on $\mathcal M_E$ uniquely determines $f$ on $\Sigma_E \times \Sigma_E$, under assumptions (\ref{fir2}), (\ref{fir5}).

\subsection{Faddeev functions}\label{sec22}

We recall also that the Faddeev solutions $\psi(x,k)$ for (\ref{fir}) satisfy the following generalized Lipmann-Schwinger integral equation
\begin{eqnarray}\label{19JanA}
\psi({x},k)=e^{ik  {x}}+ \int_{{y} \in \mathbb R^2} G({x}-{y},k) v({y}) \psi({y},k) d{y}, \label{fir13} \\
G({x},k)=g(x,k)e^{ik{x}}, \label{2DecAB} \\ g({x},k)=- \frac{1}{(2\pi)^2}
\int_{\xi \in \mathbb R^2} \frac{e^{i\xi {x}}}{|\xi|^2 + 2k \xi} d \xi,\label{2DecA}
\end{eqnarray}
where $ x \in \mathbb R^2, ~k \in \mathbb C^2\backslash \mathbb R^2, ~ k^2=E>0$; see e.g. \cite{faddeev66}, \cite{RN2}. In addition, we consider (\ref{fir13}) as an equation for $\psi=e^{ikx} \mu(x,k)$, where $\mu(\cdot,k) \in L^\infty(\mathbb R^2)$ at fixed $k$.
Note that equation (\ref{19JanA}) can be rewritten as
\begin{equation}\label{air2}
\mu({x},k)=1 + \int_{{y} \in \mathbb R^2} g({x}-{y},k) v({y}) \mu({y},k) d{y},
\end{equation}
where $ x \in \mathbb R^2, ~k \in \mathbb C^2\backslash \mathbb R^2, ~ k^2=E>0$; see e.g. \cite{RN2}.

Under assumptions (\ref{fir2}) and, in particular, under assumptions (\ref{fir9}), equations (\ref{fir13}), (\ref{air2})  are uniquely solvable for $\mu(\cdot,k) \in L^\infty(\mathbb R^2)$ at fixed $k$ if $k \in \left ( \Sigma_E \backslash \mathbb  S^1_{\sqrt{E}} \right ) \backslash \mathcal E_E,$ where $\mathcal E_E$ is the set of the Faddeev exceptional points on $\Sigma_E \backslash \mathbb S^1_{\sqrt{E}} $;  see e.g. \cite{RN2}.

Note also that, due to estimates  (3.16)-(3.18) of \cite{RN2}, the following estimates hold for some constant $c_0>0:$

\begin{eqnarray}\label{pic1}
|G^+(x,\sqrt{E})| \leq  c_0 |x|^{-1/2} E^{-1/4}, \\ \label{pic2}
|g(x,k)| \leq c_0 |x|^{-1/2} |\Re k|^{-1/2},
\end{eqnarray}
where $G^+,g$ are defined in (\ref{eq1}), (\ref{2DecAB}), $x \in \mathbb R^2$, ~$k \in \mathbb C^2  \backslash \mathbb R^2$, ~$k^2=E>0$.

In addition, under  assumptions (\ref{fir9}), as a corollary of (\ref{pic1}), (\ref{pic2}), in a similar way to  proposition 4.1  in \cite{RN2}, we have that
\begin{eqnarray}\label{air3}
\|A(k)\|_{L^\infty(\mathbb R^2) \rightarrow L^\infty(\mathbb R^2) }  \leq M(\|v\|_{L^\infty(D)}, D, E,\rho), \\ \nonumber
k \in \mathbb C^2, ~k^2=E>0, ~|\Im k|=\rho>0,
\end{eqnarray}

\begin{equation}\label{fir15}
\Sigma_{E,\rho} \cap \mathcal E_E = \emptyset ~ \mbox{ if } ~ \rho >\rho_1(\|v\|_{L^\infty(D)}, D,E), ~ E>0,
\end{equation}
where $A(k)$ is the linear integral operator of equation (\ref{air2}),
\begin{equation}\label{air4}
M(q, D, E,\rho) = \frac{c_0 q I_1(D)} {(E+\rho^2)^{1/4}},
\end{equation}
\begin{eqnarray}\label{pic3}
\rho_1 = \left [ \max ([c_0q I_1(D)]^4-E , 0) \right ]^{1/2},
\\ \nonumber
q \geq 0, \quad I_1(D) = \max_{x \in \R^2} \int_D \frac{dy}{|x-y|^{1/2}}.
\end{eqnarray}

In addition to $\psi$, we consider also the generalized Faddeev scattering amplitude $h(k,l)$ defined by the formula
\begin{equation}\label{eq1303B}
h(k,l) = \frac{1}{(2\pi)^2}  \int_{\mathbb R^2} e^{-ily} v(y) \psi(y,k)dy,
\end{equation}
where $(k,l) \in \left ( \Sigma_E \backslash \mathbb S^1_{\sqrt{E}} \right )\times \Sigma_E$; see e.g. \cite{faddeev74}, \cite{RN2}.  Here we assume also that $\Im k = \Im l$ if (\ref{fir5}) is not assumed.

Note that, under assumption (\ref{fir9}),
\begin{eqnarray}\label{fir12}
\mbox{$h$ is (complex-valued) real-analytic on $\left (\left ( \Sigma_E \backslash \mathbb S^1_{\sqrt{E}} \right )\backslash \mathcal E_E \right ) \times \Sigma_E$}, \\ \nonumber \mbox{ $h(k,\cdot)$ is holomorphic on $\Sigma_E$ at fixed $k$.}
\end{eqnarray}
We say that a complex-valued function is real-analytic if its real and imaginary parts are real-analytic.

\subsection{$\overline{\partial}$-equation on the Faddeev eigenfunctions}\label{sec23}

We recall that the following isomorphic relations are valid:
\begin{equation}\label{fir16}
\Sigma_E \approx \mathbb C \backslash 0, \quad \mathbb S^1_{\sqrt{E}} \approx T = \{ \lambda \in \mathbb C ~ : ~ |\lambda|=1 \}.
\end{equation}
More precisely:
\begin{eqnarray}\label{fir17}
k=(k_1,k_2) \in \Sigma_E \Rightarrow \lambda =\lambda(k):= \frac{k_1+ik_2}{\sqrt{E}} \in \mathbb C \backslash 0, \\ \nonumber
k=(k_1,k_2) \in \mathbb S^1_{\sqrt{E}}  \Rightarrow \lambda (k) \in \mathbb T;
\end{eqnarray}
\begin{equation}\label{fir18}
\lambda \in \mathbb C \backslash 0 \Rightarrow k=k(\lambda,E) \in \Sigma_E, \quad \lambda \in T \Rightarrow k=k(\lambda) \in \mathbb S^1_{\sqrt{E}}  ,
\end{equation}
where
\begin{equation}\label{kla}
k(\lambda,E)=(k_1(\lambda,E),k_2(\lambda,E)), ~ k_1=\left(\lambda + \frac{1}{\lambda} \right ) \frac{\sqrt{E}}{2},\quad k_2=\left( \frac{1}{\lambda} - \lambda \right ) \frac{i\sqrt{E}}{2}.
\end{equation}
Note also that
\begin{equation}\label{zhir1}
\left |\Re k(\lambda,E)\right |=\frac{\sqrt{E}}{2} \left (|\lambda|+|\lambda|^{-1} \right ), ~ \left |\Im k(\lambda,E) \right | = \frac{\sqrt{E}}{2} | \left |\lambda|-|\lambda|^{-1} \right |, ~ \lambda \in \mathbb C \backslash0, ~ E>0.
\end{equation}

Let
\begin{eqnarray}\label{mir15}
\mbox{ $L_{p,\nu}(\mathbb C)$ be the function space on $\mathbb C$ consisting of the functions $u$ such that  }\\ \nonumber
\mbox{$u,u_\nu \in L_p(\mathcal D_1)$ with the norm } \|u \|_{L_{p,\nu}} = \|u \|_{L_{p}(\mathcal D_1)} + \|u_\nu \|_{L_{p}(\mathcal D_1)},
\end{eqnarray}
where $p \geq 1, ~ \nu \geq 0$,
\begin{equation}\label{mir16}
\mbox{ $u_\nu(\lambda):= |\lambda|^{-\nu}  u(\lambda^{-1})$ },
\end{equation}
\begin{equation}\label{mir17}
\mathcal D_1 = \{ \lambda \in \mathbb C ~ : ~ |\lambda| \leq 1 \}.
\end{equation}

It is known that the function $\psi$ of subsection \ref{sec22} has, in particular, the following properties, under assumptions (\ref{fir2}) and, in particular, under assumptions (\ref{fir9}):
\begin{equation}\label{fir21}
\psi(x,k(\lambda))= e^{ik(\lambda)x} (1+o(1)), \mbox{ if } \lambda \rightarrow 0 \mbox{ or } \lambda \rightarrow \infty,
\end{equation}
\begin{eqnarray}\label{fir19}
\frac{\partial}{\partial \overline{\lambda}} \psi (x,k(\lambda))=\frac{\sgn(|\lambda|^2-1)}{\overline{\lambda}} b(k(\lambda))  \psi\left (x,k\left (-\frac{1}{\overline \lambda} \right ) \right ), \\ \nonumber
\psi\left (x,k\left (-\frac{1}{\overline \lambda} \right ) \right ) = \overline{\psi(x,(k(\lambda)))}, \quad
k(\lambda) \in \left ( \Sigma_E \backslash \mathbb  S^1_{\sqrt{E}} \right ) \backslash \mathcal E_E,
\end{eqnarray}
where $x \in \mathbb R^2, ~ k(\lambda)=k(\lambda,E)$ is defined by (\ref{kla}),
\begin{equation}\label{fir20}
b(k):= h (k, -\overline{k}),
\end{equation}
where $h$ is defined by (\ref{eq1303B}); see e.g. \cite{gm}, \cite{RN2}.

Note that $\overline{\partial}-$equations like   (\ref{fir19}) go back to \cite{ayf}, \cite{bc1985}.

\subsection{Some estimates related with $\overline{\partial}-$equation (\ref{fir19})}\label{sec24}

In particular, as a corollary of (\ref{fir21}),
\begin{equation}\label{zir6}
\mbox{$\psi(x,(k(\lambda))) \neq 0$ if
$|\lambda|$ is sufficiently small or if $|\lambda|$ is sufficiently large. }
\end{equation}
In addition, under assumptions (\ref{fir9}), as a corollary of (\ref{air3}), we have
\begin{eqnarray}\label{zhir2}
|\mu(x,k(\lambda))| \leq (1- M(q,D,E,\rho))^{-1}, ~
\\ \nonumber
x \in \mathbb R^2, ~ k(\lambda)=k(\lambda,E)\in \Sigma_{E,\rho}, ~ \rho>\rho_1(q,D,E), ~ \|v\|_{L^\infty(D)} < q, ~
\end{eqnarray}
where $M$ is defined by (\ref{air4}), $\rho_1$ is defined by (\ref{pic3}).

In connection with equation (\ref{fir19}) we consider also
\begin{equation}\label{mir14}
u_{E,\rho}(\lambda)= \frac{1}{\overline{\lambda}} \chi_{E,\rho}(k(\lambda))b(k(\lambda)),
\end{equation}
where $\chi_{E,\rho}$ is defined by (\ref{fir14B}).

Under assumptions (\ref{fir9}), we have:
\begin{equation}\label{air1}
u_{E,\rho}\in L_{p,2}(\mathbb C), \quad 2<p<4,
\end{equation}
where $\rho > \rho_1(\|v\|_{L^\infty(D)}, D,E);$
\begin{equation}\label{chgM1}
\|u_{E,\rho}\|_{L_{p,2}} \leq q c_1(D,p,E) (1-M(q,D,E,\rho))^{-1},
\end{equation}
\begin{equation}\label{air11}
\|u_{E,\rho}\|_{L_{p,2}} =  O\left ( q \right ) \mbox{ as } q \rightarrow 0,
\end{equation}
for fixed $E>0$, $\rho> \rho_1(q,D,E), D$ and $p$, where $\|v\|_{L^\infty(D)} \leq q$, $M$ is defined by (\ref{air4}), $c_1$ is a positive
 constant,  $2<p<4$;
\begin{equation}\label{air5}
|\overline{\lambda} u_{E,\rho}(\lambda) | \leq q (2\pi)^{-2} (1-M(q,D,E,\rho))^{-1} \int_D dx , ~ \lambda \in \mathbb C,
\end{equation}
where $\|v\|_{L^\infty(D)} \leq q$, $M$ is defined by (\ref{air4}), $\rho>\rho_1(\|v\|_{L^\infty(D)}, D,E)$; see formulas (4.4), (4.12), (4.18), (4.19) of \cite{RN2}.
In connection with (\ref{air1})-(\ref{air11}) we recall that  $L_{p,2}(\mathbb C)$ is defined in (\ref{mir15}).

\subsection{Final remarks}\label{sec25}
We recall also that, under the assumptions (\ref{fir22}), (\ref{fir9}), at fixed $E$, the scattering amplitude $f$ uniquely determines the Dirichlet-to-Neumann map $\Phi$ and vice versa; see proposition 4 in \cite{novikov88}.

In turn, $\Phi(E)$ uniquely determines $h$ on $\left ( \left ( \Sigma_E \backslash \mathbb  S^1_{\sqrt{E}} \right ) \backslash \mathcal E_E \right ) \times \Sigma_E$; see \cite{novikov88}.

Note also that $f$ at fixed $E$ uniquely determines $h$ on $\left ( \left ( \Sigma_E \backslash \mathbb  S^1_{\sqrt{E}} \right ) \backslash \mathcal E_E \right ) \times \Sigma_E$ via
a two-dimensional analogue of the construction given in \cite{novikov94}.

As a corollary, problems \ref{prob11}, \ref{prob12} of section 1 are reduced to problem \ref{prob31} of section 3.

\section{Global generalized Riemann-Hilbert problem}\label{sec3}
Let
\begin{eqnarray}\label{fir23}
\Lambda = \Lambda_{E,\rho} = \left \{ \lambda \in \mathbb C ~ : ~ \frac{\sqrt{E}}{2} \left | |\lambda|-|\lambda|^{-1}\right | < \rho \right \}, ~ E>0, ~ \rho > 0, \mbox{ and}\\ \nonumber
\quad \quad \mbox{ $\partial \Lambda = \partial \Lambda_{E,\rho}$ be the boundary of $\Lambda$ in $\mathbb C$ with the standard orientation.}
\end{eqnarray}
Note that
\begin{equation}\label{fir24}
\Sigma_{E,\rho} \approx \mathbb C \backslash \Lambda_{E,\rho},
\end{equation}
where this isomorphism is given by formulas (\ref{fir17}), (\ref{fir18}).

Let \[
W(\lambda,\varsigma) =  \frac{i}{2}{\rm sgn}(|\lambda |^2-1)\left [\frac{1}{\varsigma}\ln w_1(\lambda,\varsigma)+\varsigma\ln w_2(\lambda,\varsigma) \right ]
\]
\begin{equation}\label{4Jan1}
+\int_{|\eta|=1}\frac{1}{2(\varsigma-\eta)}  \theta \left  [\sgn(|\lambda|^2-1)i \left  (\frac{|\lambda|\eta}{\lambda}- \frac{\lambda}{|\lambda|\eta}\right )\right ]|d\eta|
, \quad \lambda,\varsigma \in \partial \Lambda,
\end{equation}
where
$$
w_1=\frac{\varsigma-\lambda}{\varsigma-\frac{\lambda}{|\lambda|}}, \quad w_2= \frac{\frac{-1}{\varsigma}-\overline{\lambda} } {\frac{-1}{\varsigma}-\frac{\overline{\lambda}}{|\lambda|}},
$$
$\theta$ is the standard Heaviside step function.

\begin{remark}\label{remA} Note that
\begin{equation}\label{arg}
|\arg w_i(\lambda,\varsigma)| <\pi,~~\lambda,\varsigma \in \partial \Lambda, ~~ i=1,2,
\end{equation}
and the logarithms in (\ref{4Jan1}) are well defined by the condition $|\Im \ln w_i|<\pi$.
\end{remark}
In particular, we have
\begin{equation}\label{zhir5}
W \in L_p(\partial \Lambda \times \partial \Lambda ), ~ p \geq 1, ~ \partial \Lambda = \partial \Lambda_{E,\rho}, ~ E>0, ~ \rho >0.
\end{equation}

\begin{lemma}\label{13NovA}
Let $v$ satisfy (\ref{fir9}) and let $\rho \geq \rho_1(\|v\|_{L^\infty(D)}, D,E),$  where $\rho_1$ is the constant in  (\ref{fir15}).
 Let $\psi^+,\psi$ be the eigenfunctions of subsections \ref{sec21}, \ref{sec22}. Then the following relation holds:
\begin{equation}\label{1DecC}
\psi(x,k(\lambda))= \psi^+(x,k(\lambda)) + \int_{\partial \Lambda} W(\lambda,\varsigma)h(k(\lambda),k(\varsigma))\psi^+(x,k(\varsigma))d\varsigma, \quad \lambda \in \partial \Lambda.
\end{equation}
where $k(\lambda)=k(\lambda,E)$ is given by (\ref{kla}), $W(\lambda,\varsigma)=W(\lambda,\varsigma,E) $ is given by (\ref{4Jan1}), $h$ is defined by  (\ref{eq1303B}) and the integration is taken according to the standard orientation of the $\partial \Lambda$.
\end{lemma}
Lemma \ref{13NovA} is proved in section \ref{sec5}.

Note that, under assumptions (\ref{fir9}), as a corollary of (\ref{eq1303B}), (\ref{zhir2}), we have
\begin{eqnarray}\label{zhir6}
|h(k(\lambda),k(\varsigma))| \leq q (2\pi)^{-2} e^{2\rho L} (1-M(q,D,E,\rho))^{-1} \int_D dx, ~ \\ \nonumber \lambda,\varsigma \in \partial \Lambda=\partial \Lambda_{E,\rho}, ~
\rho>\rho_1(q,D,E), \|v\|_{L^\infty(D)} \leq q,
\end{eqnarray}
where $M,\rho_1$ are defined by (\ref{air4}), (\ref{pic3}),
\begin{equation}\label{zhir7}
L=\max_{x \in \partial D} |x|.
\end{equation}

As a corollary of properties (\ref{fir8}), (\ref{fir21}), (\ref{fir19}), (\ref{1DecC}) of the functions $\psi^+$ and $\psi$ (and using (\ref{air1}), (\ref{air5})), we obtain
the following proposition:
\begin{proposition}\label{th13A}
Let $v$ satisfy (\ref{fir9}) and let $\rho \geq \rho_1(\|v\|_{L^\infty(D)}, D,E),$  where $\rho_1$ is the constant in  (\ref{fir15}).
Let $\psi,\psi^+$ be the eigenfunctions of subsections \ref{sec21}, \ref{sec22}. Then at fixed $x \in \mathbb R^2$:
\begin{enumerate}
\item
$\psi^+(x,k(\lambda))$ is holomorphic in $\lambda \in \Lambda$ and is continuous in $\lambda \in \Lambda \cup \partial \Lambda$;
\item $\psi(x,k(\lambda))$ has the properties  (\ref{fir21}), (\ref{fir19}) for  $\lambda \in ( \mathbb C \backslash 0) \backslash (\Lambda \cup \partial \Lambda) $ and is continuous in $\lambda \in ( \mathbb C \backslash 0) \backslash \Lambda  $;
\item $\psi^+,\psi$ are related on $\partial \Lambda$ via (\ref{1DecC}).
\end{enumerate}
\end{proposition}
Now we consider the following generalized inverse scattering problem for equation (\ref{fir}).

 \begin{problem}\label{prob31} Given the Faddeev functions $h$ on $\partial \Lambda \times \partial \Lambda$ and $b$ on $(\mathbb C \backslash 0) \backslash \Lambda$, find potential $v$ on $D$. \end{problem}

The approach of the present work for solving problems \ref{prob11}, \ref{prob12} and \ref{prob31} is based on the reduction of problem \ref{prob31} to the following generalized Riemann-Hilbert problem.

\begin{problem}\label{prob32} Given functions $h$ on $\partial \Lambda \times \partial \Lambda$ and $b$ on $(\mathbb C \backslash 0) \backslash \Lambda$,
find functions $\psi^+$ on $\Lambda$ and $\psi$ on $(\mathbb C \backslash 0) \backslash \Lambda$ satisfying the properties of the items 1,2,3 of proposition
\ref{th13A}.
\end{problem}

Note that in problems \ref{prob31}, \ref{prob32} we consider $h,b$ and $\psi^+,\psi$ as
\begin{equation}\label{smir5}
h=h(\lambda,\zeta,E)=h(k(\lambda),k(\zeta)),  ~ \lambda,\zeta \in \partial \Lambda,
\end{equation}
\begin{equation}\label{smir6}
\psi^+=\psi^+(x,\lambda,E)=\psi^+(x,k(\lambda)),  ~ \lambda \in \Lambda,
\end{equation}
\begin{equation}\label{smir7}
\psi=\psi(x,\lambda,E)=\psi(x,k(\lambda)), ~ b=b(\lambda,E)=b(k(\lambda)), ~ \lambda \in ( \mathbb C \backslash 0) \backslash \Lambda,
\end{equation}
where $k(\lambda)=k(\lambda,E)$ is defined by (\ref{kla}), $\Lambda=\Lambda_{E,\rho}$, $\partial \Lambda = \partial \Lambda_{E,\rho}$  are defined in (\ref{fir23}), $h$ is defined by (\ref{eq1303B}) and
$b$ is defined by (\ref{fir20}).

In addition, if $\psi$ is the function of subsections \ref{sec22}, \ref{sec23}, \ref{sec24}, then it determines the potential easily. Indeed, due to (\ref{fir}), (\ref{zir6}), we have
\begin{eqnarray}\label{aprilA1}
v(x)=\frac{(\Delta_x+E)\psi(x,k(\lambda))}{\psi(x,k(\lambda))} \mbox{ for all } x \in \mathbb R^2 \\ \nonumber
\mbox{if $|\lambda|$ is sufficiently small or  if $|\lambda|$ is sufficiently large. }
\end{eqnarray}

Prototypes of problems \ref{prob31}, \ref{prob32} for the case of equation (\ref{fir})  with $E<0$ were considered in section 8 of \cite{RN2}.

Generalized Riemann-Hilbert problems like problem \ref{prob32} go back to \cite{manakov81} and to \cite{fa83}, \cite{gn85}, \cite{gm}.

We say that problem \ref{prob32} is a generalized Riemann-Hilbert-Manakov problem.

We say that the results of lemma \ref{13NovA} and proposition \ref{th13A} are global and that the related problem \ref{prob32} is global, since these results and problem are formulated for general $v$ satisfying (\ref{fir9}) and, in particular, without the assumption that $\mathcal E_E=\emptyset,$ where $\mathcal E_E$ is the set of Faddeev exceptional points at fixed $E$.

The reduction of problem \ref{prob31}  to problem \ref{prob32} follows from proposition \ref{th13A} and, for example, from formula (\ref{aprilA1}).

\section{Integral equations for solving problem 3.2}\label{sec4}
\subsection{Formulas and equations}\label{sec41}
Let
\begin{equation}\label{mir1}
\mu^+(\lambda):=e^{-ik(\lambda)x} \psi^+(x,\lambda,E),  \quad \lambda \in \Lambda,
\end{equation}
\begin{equation}\label{mir2}
\mu(\lambda):=e^{-ik(\lambda)x} \psi(x,\lambda,E),  \quad \lambda \in \mathbb C  \backslash \Lambda,
\end{equation}
\begin{eqnarray}\label{mir3}
r(x,\lambda,E) =e^{i(-k(\lambda)+k(-1/\overline{\lambda}))x} \frac{\sgn(|\lambda|^2-1)}{\overline{\lambda}} \chi(\lambda) b(\lambda,E) = \\ \nonumber
 e^{-2i \Re k(\lambda)  x} u(\lambda) , \quad  \lambda \in \mathbb C  \backslash 0 ,
\end{eqnarray}
\begin{equation}\label{mir4}
R(x,\lambda,\zeta,E) =e^{i(k(\zeta)-k(\lambda))x} W(\lambda,\zeta,E) h(\lambda,\zeta,E), ~ \lambda,\zeta \in \partial \Lambda,
\end{equation}
where $\psi^+,\psi$ and $h=h(\lambda,\varsigma,E)=h(k(\lambda),k(\varsigma))$, $b=b(\lambda,E)=b(k(\lambda))$ are the functions of Problem \ref{prob32}, $\chi(\lambda)=\chi_{E,\rho}(k(\lambda))$ is defined via (\ref{fir14B}), $u(\lambda)=u_{E,\rho}(\lambda)$ is defined via (\ref{mir14}), $k(\lambda)=k(\lambda,E)$ is defined by (\ref{kla}), $W$ is given by (\ref{4Jan1}), $\Lambda=\Lambda_{E,\rho}$ is defined by (\ref{fir23}).

Let
\begin{equation}\label{mir5}
e(\lambda)=e(x,\lambda,E), \quad  X_j(\lambda,\zeta)=X_j(x,\lambda,\zeta,E), ~j=1,2, \quad \lambda,\zeta \in \mathbb C,
\end{equation}
be defined as the solutions of the following linear integral equations:
\begin{equation}\label{mir6}
e(\lambda)=1- \frac{1}{\pi} \int_{\mathbb C} r(x,\zeta,E) \overline{e(\zeta)} \frac{d\Re \zeta d\Im \zeta}{\zeta- \lambda},
\end{equation}
\begin{equation}\label{mir7}
X_1(\lambda,\zeta) + \frac{1}{\pi} \int_{\mathbb C} r(x,\eta,E) \overline{X_1(\eta,\zeta)}\frac{d\Re \zeta d\Im \zeta} {\eta - \lambda}= \frac{1}{2(\zeta- \lambda)},
\end{equation}
\begin{equation}\label{mir8}
X_2(\lambda,\zeta) + \frac{1}{\pi} \int_{\mathbb C} r(x,\eta,E) \overline{X_2(\eta,\zeta)}\frac{d\Re \zeta d\Im \zeta} {\eta - \lambda}= \frac{1}{2i(\zeta- \lambda)}.
\end{equation}

In addition, we consider also
\begin{equation}\label{mir9}
\Omega_1(\lambda,\zeta) := X_1(\lambda,\zeta) + i X_2(\lambda,\zeta), \quad \Omega_2(\lambda,\zeta) := X_1(\lambda,\zeta) - i X_2(\lambda,\zeta), \quad \lambda,\zeta \in \mathbb C.
\end{equation}

Note that if (\ref{air1}) is fulfilled, then equation (\ref{mir6}) for $e(\cdot)$ and equations (\ref{mir7}), (\ref{mir8})
for $X_j(\cdot,\zeta), ~j=1,2,$ are uniquely solvable in $L_{q,0}(\mathbb C), ~ p/(p-1) \leq q < 2$, where $L_{p,\nu}$ is defined in (\ref{mir15}). In addition:
\begin{eqnarray}\label{fr5}
e(\cdot) \in C(\mathbb C \cup \infty), \quad e(\infty)=1, \\ \nonumber
|e(\lambda)-1| \leq c_2(r_0,p),
\end{eqnarray}
\begin{equation}\label{zir1}
\left | \Omega_1(\lambda,\zeta) - \frac{1}{\zeta - \lambda} \right | < c_2(r_0,p) \frac{1}{2|\zeta - \lambda|^{2/p}},
\end{equation}
\begin{equation}\label{zir2}
\left | \Omega_2(\lambda,\zeta) \right | < c_2(r_0,p) \frac{1}{2|\zeta - \lambda|^{2/p}},
\end{equation}
where
\begin{equation}\label{zir3}
r_0 =  \| r(x,\cdot,E) \|_{L_{p,2}} , \quad \lim_{r_0 \rightarrow 0} c_2(r_0,p)=0.
\end{equation}
Note that $r_0$ is independent of $x \in \mathbb R^2$.

In connection with the functions $e,X_1,X_2,\Omega_1,\Omega_2$ and related results we refer to chapter 3 of \cite{vekua} and to section 6 of \cite{RN2}.

We define
\begin{equation}\label{aprilA2}
\psi'(\lambda) = \left \{\begin{array}{l}
\psi^+(\lambda), ~ \lambda \in \Lambda \cup \partial \Lambda, \\
\psi(\lambda), ~ \lambda \in (\mathbb C \backslash 0) \backslash  \Lambda, \\
\end{array} \right .
\end{equation}
where $\psi^+,\psi$ are the functions of problem \ref{prob32}. In addition, we consider $\mu',\mu^+,\mu$, where
\begin{equation}\label{aprilA3}
\psi'(\lambda) = e^{ik(\lambda)x} \mu'(\lambda) = e^{ik(\lambda)x}\left \{\begin{array}{l}
\mu^+(\lambda), ~ \lambda \in \Lambda \cup \partial \Lambda, \\
\mu(\lambda), ~ \lambda \in (\mathbb C \backslash 0) \backslash  \Lambda. \\
\end{array} \right .
\end{equation}

\begin{theorem}\label{thmir1}

Let the data $h$ and $b$ of problem \ref{prob32} satisfy the following conditions:
\begin{equation}\label{zir7}
\mbox{ $u_{E,\rho} \in L_{p,2}(\mathbb C), ~ 2<p<4$, }
 \end{equation}
\begin{equation}\label{zir8}
 \mbox{$h(\cdot,\cdot,E) \in C(\partial \Lambda \times \partial \Lambda)$, }
 \end{equation}
where $u_{E,\rho}$ is defined by (\ref{mir14}), $W$ is defined by (\ref{4Jan1}), $\partial \Lambda=\partial \Lambda_{E,\rho}$ is defined in (\ref{fir23}), $\rho >0$. Let $\psi'$ be a solution of problem \ref{prob32}.
Then for $\mu'$  defined by (\ref{aprilA3}) the following formula holds:
\begin{equation}\label{mir11}
\mu'(\lambda)= e(\lambda) +\frac{1}{2\pi i } \int_{\partial \Lambda} \Omega_1(\lambda,\zeta) K(\zeta) d\zeta -\frac{1}{2\pi i } \int_{\partial \Lambda} \Omega_2(\lambda,\zeta) \overline{K(\zeta)} d\overline{\zeta}, \quad \lambda \in \mathbb C \backslash \partial \Lambda,
\end{equation}
where the integration is taken according to the standard orientation of $\partial \Lambda$,
\begin{equation}\label{mir12}
K(\lambda):=\mu^+(\lambda)-\mu(\lambda), \quad \lambda \in \partial \Lambda.
\end{equation}
In addition, this $K=K(x,\lambda,E)$ satisfies the following linear integral equation
\begin{equation}\label{mir13}
\begin{array}{l}
K(\lambda) + \left . \int_{\partial \Lambda} R(x,\lambda,\lambda',E) \right ( e(\lambda') +  \\
\left .  \frac{1}{2\pi i } \int_{\partial \Lambda} \Omega_1(\lambda'(1-0(|\lambda'|-1)),\zeta) K(\zeta)d\zeta - \frac{1}{2\pi i } \int_{\partial \Lambda} \Omega_2(\lambda',\zeta) \overline{K(\zeta)} d\overline{\zeta }\right ) d\lambda'=0, \quad \lambda \in \partial \Lambda,
\end{array}
\end{equation}
where $R$ is defined by (\ref{mir4}), $\Omega_1,\Omega_2$ are defined by (\ref{mir9}) and the integrations are taken according to the standard orientation of $\partial \Lambda$.
\end{theorem}

Note also that
\begin{eqnarray}\label{zir4}
\int_{\partial \Lambda} \Omega_1(\lambda'(1-0(|\lambda'|-1)),\zeta) K(\zeta)d\zeta = \\ \nonumber
\lim_{0<\varepsilon \rightarrow 0} \int_{\partial \Lambda} \Omega_1(\lambda'(1-\varepsilon(|\lambda'|-1)),\zeta) K(\zeta)d\zeta, ~ \lambda' \in \partial \Lambda.
\end{eqnarray}

Formula (\ref{mir11}) is similar to formula (6.7) of \cite{RN2}. Equation (\ref{mir13}) is similar to equation (6.11) of \cite{RN2}.

Theorem \ref{thmir1} is proved in section \ref{sec6}.

Consider
\begin{eqnarray}\label{frl1}
I(\lambda) = I(x,\lambda,E) = -\int_{\partial \Lambda} R(x,\lambda,\lambda',E)  e(\lambda')d \lambda ', \quad \lambda \in \partial \Lambda, \\ \label{frl1B}
A_1(\lambda,\zeta)= A_1(x,\lambda,\zeta,E) =  \frac{1}{2\pi i } \int_{\partial \Lambda} R(x,\lambda,\lambda',E)  \Omega_1(\lambda'(1-0(|\lambda'|-1)),\zeta)
d\lambda ', \\ \nonumber
A_2(\lambda,\zeta)= A_2(x,\lambda,\zeta,E) =  \frac{-1}{2\pi i } \int_{\partial \Lambda} R(x,\lambda,\lambda',E)  \Omega_2(\lambda',\zeta)
d\lambda ', \quad \lambda,\zeta \in \partial \Lambda,
\end{eqnarray}
where $R,e, \Omega_1,\Omega_2$ are the functions of (\ref{mir4}), (\ref{mir5}), (\ref{mir9}).

\begin{proposition}\label{prop41} Let the assumptions of theorem \ref{thmir1} be fulfilled and $K$ be the function of (\ref{mir12}), (\ref{mir13}). Then $K,\overline{K}$ satisfy the following system of linear integral equations
\begin{eqnarray}\label{frl2}
K(\lambda) + \int_{\partial \Lambda} A_1(\lambda,\zeta) K(\zeta) d \zeta + \int_{\partial \Lambda} A_2(\lambda,\zeta) \overline{K(\zeta)} d \overline{\zeta} = I(\lambda),  ~ \lambda \in \partial \Lambda, \\ \label{fr12B}
\overline{K(\lambda)} + \int_{\partial \Lambda} \overline{A_2(\lambda,\zeta)} K(\zeta) d \zeta + \int_{\partial \Lambda} \overline{A_1(\lambda,\zeta) } ~ \overline{K(\zeta)} d \overline{\zeta} = \overline{I(\lambda)}, ~ \lambda \in \partial \Lambda,
\end{eqnarray}
where $I,A_1,A_2$ are defined by (\ref{frl1}), (\ref{frl1B}). In addition,
\begin{eqnarray}\label{jun24A}
I \in L_2(\partial \Lambda), \quad A_j \in L_2(\partial \Lambda \times \partial \Lambda ), ~ j=1,2, \\ \label{Aug2A}
\|A_j\|_{L_2} \rightarrow 0 ~ \mbox{ for } \|h\|_{C} \rightarrow 0, \quad r_0 \leq r_{\rm{fixed}}, ~ j=1,2,
\end{eqnarray}
where $|x|<c$ for fixed $c>0$,  $r_0$ is defined in (\ref{zir3})
\end{proposition}

Proposition \ref{prop41} is proved in section 6.

\subsection{Analysis of equations}\label{subsec42}
Due to estimates (\ref{jun24A}), the system (\ref{frl2}), (\ref{fr12B}) can be considered as a Fredholm linear integral equation of the second type for the vector-function
$(K,\overline{K}) \in L_2(\partial \Lambda,\mathbb C^2)$ with parameters $x \in \mathbb R^2$ and $E>0$.

The modified Fredholm determinant $\rm{det} A$ for system (\ref{frl2}), (\ref{fr12B}) can be defined by means of the formula:
\begin{equation}\label{chg2A}
\ln \rm{det} A = Tr(\ln (Id + A)-A),
\end{equation}
where system (\ref{frl2}), (\ref{fr12B})  is written as
\begin{equation}\label{chg2B}
(Id+A) \left ( \begin{array}{c} K \\ \overline{K} \end{array} \right ) = \left ( \begin{array}{c} I \\ \overline{I} \end{array} \right ).
\end{equation}
For the precise definition of $\rm{det} A$, see \cite{gokhbergkrein}.

In addition, we have the following lemmas:
\begin{lemma}\label{lemmachg1}
Let $v$ satisfy (\ref{fir9}) for fixed $D$ and $\Lambda=\Lambda_{E,\rho}$ be defined by (\ref{fir23}) for fixed $E$ and $\rho$. Let $A_1,A_2,I$ correspond to $v$ according to formulas (\ref{19JanA})-(\ref{air2}), (\ref{eq1303B}), (\ref{fir20}), (\ref{smir5}), (\ref{smir7}), (\ref{mir3}), (\ref{mir4}), (\ref{frl1}), (\ref{frl1B}). Let $|x|<c$ for fixed $c>0$.  Then:
\begin{equation}\label{chg2C}
\|A_j\|_{L^2(\partial \Lambda \times \partial \Lambda)} \rightarrow 0, ~ \|I\|_{L^2(\partial \Lambda)} \rightarrow 0, ~ \mbox{ for } \|v\|_{L^\infty(D)} \rightarrow 0, ~ j=1,2;
\end{equation}
system (\ref{frl2}), (\ref{fr12B})   for $(K,\overline{K}) \in L_2(\partial \Lambda,\mathbb C^2)$ is uniquely solvable by the method of successive approximations when $\|v\|_{L^\infty(D)}$ is sufficiently small (for fixed $D,E$,$\rho$ and $c$).
\end{lemma}

Actually, lemma \ref{lemmachg1} follows from estimates (\ref{air11}), (\ref{zhir6}), (\ref{fr5})-(\ref{zir2}), (\ref{Aug2A}).

\begin{lemma}\label{fried1}
Let $v$ satisfy (\ref{fir9}) for fixed $D$ and $\Lambda=\Lambda_{E,\rho}$ be defined by (\ref{fir23}) for fixed $E$ and $\rho$, where $\rho>\rho_1(q,D,E)$,
$\|v\|_{L^\infty(D)} < q$, $\rho_1$ is defined by (\ref{pic3}).
Let $A_1,A_2,I$ correspond to  $sv$  according to formulas (\ref{19JanA})-(\ref{air2}), (\ref{eq1303B}), (\ref{fir20}), (\ref{smir5}), (\ref{smir7}), (\ref{mir3}), (\ref{mir4}), (\ref{frl1}), (\ref{frl1B}) (with $sv$ in place of $v$), where $s \in ]-s_1,s_1[,$ where $s_1={q}/{\|v\|_{L^\infty(D)}}$.
And let $\rm{det} A =\rm{det} A(x,s), ~ x \in \mathbb R^2, s \in ]-s_1,s_1[,$ be the modified Fredholm determinant of the related system (\ref{frl2}), (\ref{fr12B}) (where $\rm{det} A$ depends also on $v,E$ and $\rho$).
Then:
\begin{equation}\label{fried2}
\rm{det} A(x,0)=1, \quad x \in \mathbb R^2,
\end{equation}
\begin{equation}\label{fried3}
\rm{det} A \in C(\mathbb R^2 \times ]\!-s_1,s_1[,\mathbb C ),
\end{equation}
\begin{equation}\label{fried4}
\rm{det} A(x,\cdot) \mbox{ is real-analytic on }   ]\!-s_1,s_1[ \mbox{ for fixed } x \in \mathbb R^2.
\end{equation}
\end{lemma}
Lemma \ref{fried1} is proved in section \ref{sec7}.

Using lemma \ref{fried1} we obtain, in particular, the following result:

\begin{proposition}\label{nchg1}
Let $\Lambda=\Lambda_{E,\rho}$ be defined by (3.1) for fixed $E$ and $\rho$, where $\rho>\rho_1(q,D,E)$, $\rho_1$ is defined by (\ref{pic3}), $D$ is a fixed open bounded domain in $\mathbb R^2$, $q$ is a fixed positive number. Then for almost each $v$ satisfying (\ref{fir9}) with $\|v\|_{L^{\infty}(D)} \leq q$ the system (\ref{frl2}), (\ref{fr12B})  corresponding to $v$ (according to formulas (\ref{eq1303B}), (\ref{fir20}), (\ref{smir5}), (\ref{smir7}), (\ref{mir3}), (\ref{mir4}), (\ref{frl1}), (\ref{frl1B})) is uniquely solvable for almost each $x \in \mathbb R^2$.
\end{proposition}

\begin{remark}\label{nchg2}
We understand the statement of proposition \ref{nchg1} in the sense that if $v$ satisfies (\ref{fir9}) and $\|v\|_{L^{\infty}(D)} =q_1$ for fixed $q_1$, where $0<q_1<q$, then for almost each $s \in  ]-s_1,s_1[,$ where $s_1=q/q_1$, the system (\ref{frl2}), (\ref{fr12B})  corresponding to $sv$ is uniquely solvable for almost each $x \in \mathbb R^2$.
\end{remark}

\begin{remark}\label{nchg3}
If the assumptions of proposition \ref{nchg1} are fulfilled, $\|v\|_{L^{\infty}(D)} < q$, and system (\ref{frl2}), (\ref{fr12B}) corresponds to $v$, then, as a  corollary of (\ref{fried3}), the set of $x$, where the system  (\ref{frl2}), (\ref{fr12B}) is uniquely solvable, is an open set in $\mathbb R^2$.
\end{remark}

Proposition \ref{nchg1} is proved in section \ref{sec7}.

\section{Applications to the Novikov-Veselov equation}\label{secnov5}
In this section we suppose that $v$ and $\rho$ satisfy the assumptions of lemma \ref{fried1} for fixed $D$, $E$ and $q$.

We define
\begin{eqnarray}\label{nv1}
f_{s}(k,l,t) = f_{s}(k,l) \exp[ 2i t(k_1^3-3k_1 k_2^2 - l_1^3 + 3l_1l_2^2 )], ~ (k,l) \in \mathcal M_E, \\ \nonumber
  h_{s}(k,l,t) = h_{s}(k,l) \exp[ 2i t(k_1^3-3k_1 k_2^2 - l_1^3 + 3l_1l_2^2 )], ~ (k,l) \in \partial \Sigma_{E,\rho} \times  \partial \Sigma_{E,\rho}, \\ \nonumber
b_s (k,t) = b_s(k) \exp [ 2it ( k_1^3 +\overline{k}_1^3 - 3k_1k_2^2 -3 \overline{k}_1 \overline{k}_2^2 )], ~ k \in \Sigma_{E,\rho},
\end{eqnarray}
where $t \in \mathbb R$, $s \in ]-s_1,s_1[$, $s_1$ is defined as in lemma \ref{fried1} and $f_s,h_s,b_s$ are defined according to (\ref{eq1}), (\ref{eq1303}), (\ref{19JanA})-(\ref{air2}), (\ref{eq1303B}), (\ref{fir20}) with $sv$ in place of $v$. In addition:
\begin{eqnarray}\label{nv2}
h_s(k(\lambda),k(\varsigma),t) = h_s (k(\lambda),k(\varsigma))  \exp[iE^{3/2} t (\lambda^3 + \lambda^{-3}- \varsigma^3- \varsigma^{-3})] \\ \nonumber
 =: h_{s,t}(\lambda,\varsigma,E), ~ (\lambda,\varsigma) \in \partial \Lambda \times \partial \Lambda,  \\ \nonumber
b_s (k(\lambda),t)= b_s(k(\lambda))\exp[it E^{3/2} (\lambda^3+\lambda^{-3} + \overline{\lambda}^{~\!3}+\overline{\lambda}^{\!~-3})]\\ \nonumber =: b_{s,t}(\lambda,E),~ \lambda \in (\mathbb C \backslash 0) \backslash \Lambda,
\end{eqnarray}
where $t \in \mathbb R$, $s \in ]-s_1,s_1[$, $k(\lambda)=k(\lambda,E)$ is defined by (\ref{kla}), $\Lambda=\Lambda(E,\rho)$ is defined by (2.31).

We consider problem \ref{prob32} of section \ref{sec3} with $h=h_{s,t}$, $b=b_{s,t}$, $\psi^+=\psi^+_{s,t}$. As in section \ref{sec41}, we consider the reduction of this  generalized Riemann-Hilbert-Manakov problem to formulas (\ref{aprilA2}), (\ref{aprilA3}), (\ref{mir11}), (\ref{mir12}) and the system of equations (\ref{frl2}), (\ref{fr12B}), where $\mu'=\mu'_{s,t}$, $e=e_{s,t}$, $\Omega_j=\Omega_{j,s,t}, j=1,2$, $K=K_{s,t}$, $I=I_{s,t}$, $A_j=A_{j,s,t}, j=1,2$. In addition, as in section \ref{subsec42}, we consider $\det A(x,s,t)$ for the aforementioned system (\ref{frl2}), (\ref{fr12B}).

We expect that using ideas of \cite{gn85}, \cite{gn86}, \cite{gm}, \cite{RN2} and of the present work one can obtain the following result:

Suppose that $\det A(x,s,t) \neq 0$ for $x\in \mathcal X, t \in \mathcal T$ at fixed $s \in ]-s_1,s_1[$, where $\mathcal X$ is an open domain in $\mathbb R^2$, $\mathcal T$ is an open interval in $\mathbb R$, $0 \in \mathcal T$,  $s_1$ is defined in lemma \ref{fried1}. Then there is a real valued $v_{s}(\cdot,t)$ such that:
\begin{equation}
v_{s}(\cdot,0) = sv,
\end{equation}
where $sv$ is  the potential of lemma \ref{fried1};
\begin{equation}
-\Delta_x \psi^+_{s,t} + v_s(x,t) \psi^+_{s,t} = E\psi^+_{s,t}, \quad
-\Delta_x \psi_{s,t} + v_s(x,t) \psi_{s,t} = E\psi_{s,t},~ (x,t) \in \mathcal X \times \mathcal T,
\end{equation}
where $\psi^+_{s,t}=\psi^+_{s,t}(x,\lambda), ~ \lambda\in \Lambda, $ and $\psi_{s,t}=\psi_{s,t}(x,\lambda), ~ \lambda \in (\mathbb C \backslash 0) \backslash \Lambda$, solve the aforementioned problem \ref{prob32};

$v=v_{s}(x,t)$ solves the Novikov-Veselov equation (\ref{zir5}) in $\mathcal X \times \mathcal T$ with appropriate $w=w_{s}(x,t)$ (and satisfies (5.3) on $\mathcal X $).

These studies will be given in detail elsewhere.

Note that, actually, the zeroes of   $\det A(x,s,t)$ describe the blow-up points of the potential  $v_s(x,t)$.

It remains to note that in similar way to proposition \ref{nchg1} and remarks \ref{nchg2},\ref{nchg3}, for almost each $s \in ]-s_1,s_1[$, we have that $\det A(x,s,t) \neq 0$ for almost each $(x,t) \in \mathbb R^2 \times \mathbb R$; and the nonzero set of $\det \!A$ is open.

\section{Proof of lemma \ref{13NovA}}\label{sec5}
\subsection{Lemma for Green functions}\label{511}
Let
\begin{equation}\label{sic4}
z=x_1+ix_2, ~ \overline{z}=x_1- i x_2 ~ \mbox{for } x=(x_1,x_2) \in \mathbb R^2.
\end{equation}
\begin{lemma}\label{1DecLA}
The following formula holds:
\begin{equation}\label{13NovB}
G(x,k(\lambda))-G^+(x,\sqrt{E}) =\frac{1}{(2\pi)^2} \int_{\partial \Lambda} W(\lambda,\varsigma, E) e^{i\sqrt{E}/2 (\varsigma\overline{z} + z / \varsigma)}d\varsigma, \quad~ \lambda,\varsigma \in \partial \Lambda,
\end{equation}
where $G,G^+$ are defined in (\ref{2DecAB}), (\ref{eq1}), $W$ is defined by (\ref{4Jan1}), $k(\lambda)=k(\lambda,E)$ is defined in (\ref{kla}), $\Lambda=\Lambda_{E,\rho}$ is defined in (\ref{fir23}).
\end{lemma}
{\it Proof of lemma \ref{1DecLA}.} We recall that
\begin{equation}\label{sic1}
\frac{\partial}{\partial \overline{\lambda}} G(z,k(\lambda)) = \frac{\sgn (|\lambda |^2-1)}{4\pi \overline{\lambda}}
e^{ik(-1/\overline{\lambda})x}, \quad \lambda\in (\mathbb C \backslash 0)\backslash T,
 \end{equation}
\begin{equation}\label{sic2}
\frac{\partial}{\partial {\lambda}} G(z,k(\lambda)) = \frac{\sgn (|\lambda |^2-1)}{4\pi {\lambda}}
e^{ik({\lambda})x}, \quad \lambda \in (\mathbb C \backslash 0)\backslash T,
 \end{equation}
where $G$ is defined by (\ref{2DecAB}), (\ref{2DecA}), $k(\lambda)=k(\lambda,E)$ is defined by (\ref{kla}), $T$ is defined by (\ref{fir16}); see \cite{RN2}.

Note that
\begin{equation}\label{sic3}
k(-1/\overline{\lambda})x=-\frac{\sqrt{E}}{2} (\overline{\lambda} {z} + \overline{z} / \overline{\lambda}),  \quad k(\lambda)x =\frac{\sqrt{E}}{2} ({\lambda} \overline{z} + {z} / {\lambda}).
\end{equation}

Using the Cauchy formula for  $e^{ik(-1/\overline{\lambda})x}/\overline{\lambda}$ and $e^{ik({\lambda})x}/\lambda$  we have
\begin{equation}\label{sic5}
e^{ik(-1/\overline{\lambda})x}/\overline{\lambda} = \frac{1}{2\pi i} \int_{\partial \Lambda} \frac{1}{ \overline{\varsigma}}
e^{-i\sqrt{E}/2 (\overline{\varsigma} {z} + \overline{z} / \overline{\varsigma})}\frac{d\overline{\varsigma}}{\overline{\varsigma}-\overline{\lambda}},  \quad \lambda\in \Lambda,
\end{equation}
\begin{equation}\label{sic6}
e^{ik({\lambda})x}/{\lambda} = \frac{1}{2\pi i} \int_{\partial \Lambda} \frac{1}{ {\varsigma}}
e^{i\sqrt{E}/2 ({\varsigma} \overline{z} + {z} /{\varsigma})}\frac{d{\varsigma}}{{\varsigma}-{\lambda}},  \quad \lambda\in \Lambda.
\end{equation}
Due to (\ref{sic1}), (\ref{sic2}) and (\ref{sic5}), (\ref{sic6}) we have

\begin{equation}\label{sic7}
\frac{\partial}{\partial \overline{\lambda}} G(x,k(\lambda)) = \sgn (|\lambda |^2-1)\frac{-1}{2\pi i} \int_{\partial \Lambda} \frac{1}{4\pi \overline{\varsigma}}
e^{-i\sqrt{E}/2 (\overline{\varsigma} {z} + \overline{z} / \overline{\varsigma})}\frac{d\overline{\varsigma}}{\overline{\varsigma}-\overline{\lambda}},  \quad \lambda\in \Lambda, ~~|\lambda|\neq 1,
\end{equation}
\begin{equation}\label{sic8}
\frac{\partial}{\partial \lambda} G(x,k(\lambda)) = \sgn (|\lambda |^2-1)\frac{1}{2\pi i} \int_{\partial \Lambda} \frac{1}{4\pi \varsigma}
e^{i\sqrt{E}/2 ({\varsigma} \overline{z} + {z} / \varsigma)}\frac{d\varsigma}{\varsigma-\lambda}, \quad  \lambda\in \Lambda,~~|\lambda|\neq 1.
 \end{equation}
 Formulas (\ref{sic7}), (\ref{sic8}) remain also valid with $G(x,k(\lambda))$ replaced in the left hand side by $G(x,k(\lambda)) - G^+(x,\sqrt{E})$, where $G^+$ is defined in (\ref{eq1}).

 Integrating the differential equation for $G-G^+$ we obtain
 \begin{eqnarray}\label{rec}
G(x,k(\lambda)) - G^+(x,\sqrt{E})=u(z,\lambda)+\left [G(x,k(\lambda_0))-G^+(x,\sqrt{E})-u(z,\lambda_0) \right],  \\ \nonumber
\mbox{for } \lambda_0=\lambda_0(\lambda),\lambda \in \Lambda \cap \mathcal D_1, \mbox{ or for } \lambda_0(\lambda),\lambda \in \Lambda \cap (\mathbb C \backslash \mathcal D_1),
 \end{eqnarray}
where $\mathcal D_1$ is defined by (\ref{mir17}), $\lambda_0=\lambda_0(\lambda)= \frac{\lambda}{|\lambda|}(1+ 0 (|\lambda |^2-1))$,
\begin{eqnarray}\label{hic2}
u(z,\lambda)=\frac{\sgn (|\lambda |^2-1)}{2\pi i} \int_{\partial \Lambda} \frac{1}{4\pi \overline{\varsigma}}
e^{-i\sqrt{E}/2 (\overline{\varsigma} {z} + \overline{z} / \overline{\varsigma})}\ln(\overline{\varsigma}-\overline{\lambda})d\overline{\varsigma}
\\ \nonumber
-\frac{\sgn (|\lambda |^2-1)}{2\pi i} \int_{\partial \Lambda} \frac{1}{4\pi \varsigma}
e^{i\sqrt{E}/2 ({\varsigma} \overline{z} + {z} / \varsigma)}\ln(\varsigma-\lambda)d\varsigma, ~ \lambda\in \Lambda\backslash T ,
\end{eqnarray}
where notation $1+0 (|\lambda |^2-1)$ is like in (\ref{zir4}).  In the last expression logarithm is chosen such that $|\Im \ln(\cdot)|<\pi$.

We change the variable $\varsigma\to {-1}/{\overline{\varsigma}}$ in the first integral on the right and obtain the formula
\begin{equation}\label{pott}
u(z,\lambda)=-\frac{\sgn (|\lambda |^2-1)}{8\pi^2 i} \int_{\partial \Lambda}
e^{i\sqrt{E}/2 ({\varsigma} \overline{z} + {z} / \varsigma)}\left [\frac{1}{ \varsigma}\ln\left (\varsigma-\lambda \right )+\varsigma\ln\left (\frac{-1}{\varsigma}-\overline{\lambda}\right )\right]d\varsigma, \quad \lambda\in \Lambda \backslash T.
\end{equation}
In the last expression logarithm is chosen such that $|\Im \ln(\cdot)|<\pi$.

We choose $\lambda_0$ as $\lambda_0=\frac{\lambda}{|\lambda|}(1\pm 0)$ since the limiting values of  $G-G^+$ on the unit circle $T$ are given by (see \cite[section 3]{RN2}):
\begin{equation}\label{hic3}
 G(x,k(\lambda_0))-G^+(x,\sqrt{E}) = \frac{\pi i}{(2\pi)^2}  \int_{T} e^{i\sqrt{E}/2 (\varsigma\overline{z} + z / \varsigma)} \theta \left  [\sgn(|\lambda|^2-1) i \left   (\frac{|\lambda|\varsigma}{\lambda}- \frac{\lambda}{|\lambda|\varsigma}\right )\right ]|d \varsigma|,
 \end{equation}
 where $\theta$ is the Heaviside step function. Using the Cauchy formula for $e^{i\sqrt{E}/2 (\varsigma\overline{z} + z / \varsigma)} $  in (\ref{hic3}), we can rewrite (\ref{hic3}) as follows:
\begin{eqnarray}\label{hic4}
G(x,\lambda_0)-G^+(x,\sqrt{E})=
\frac{1}{8\pi^2}\int_{\varsigma_1 \in T} \left( \int_{\partial \Lambda} \frac{ e^{i\sqrt{E}/2 (\varsigma\overline{z} + z / \varsigma)} d\varsigma}{\varsigma-\varsigma_1} \right ) \times \\ \nonumber \theta \left  [\sgn(|\lambda|^2-1)i \left  (\frac{|\lambda|\varsigma_1}{\lambda}- \frac{\lambda}{|\lambda|\varsigma_1}\right )\right ]|d\varsigma_1|.
 \end{eqnarray}
 In order to complete the proof of lemma \ref{1DecLA} it remains only to put (\ref{hic4}), (\ref{pott}) into (\ref{rec}).

In addition, to justify remark \ref{remA}, we need to prove (\ref{arg}). Assume that $\varsigma$
 belongs to the part  $|\varsigma|=C$ of $\partial \Lambda =\partial \Lambda_{E,\rho}$ where
 $$
 C=\rho / \sqrt{E}+\sqrt{(\rho/\sqrt{E})^2+1}.
 $$
 Since the point $\lambda$ belongs to the disk $|\varsigma|\leq C$ and the point $\lambda_0$ is strictly inside of the disk, the angle $\alpha$ between vectors $\varsigma-\lambda$ and $\varsigma-\lambda_0$ is strictly less then $\pi$. Thus $|\arg w_1|=|\alpha| <\pi$ in this case. If $\varsigma$
 belongs to the part  $|\varsigma|=1/C$ of the boundary of $\partial \Lambda$, then points $\lambda$ and $\lambda_0$ belong to the part of the ray (emitted from $\lambda=0$) through the point $\lambda$. This part belongs to the region $|\varsigma|\geq 1/C$, and $|\arg w_1|=|\alpha| <\pi/2$ in this case. After the estimate  (\ref{arg}) for $w_1$ is proved, the estimate for $w_2$ becomes obvious if we replace $-1/\varsigma$ by $\overline{\varsigma}$.

\qed
\subsection{Final part of proof of lemma \ref{13NovA}}
Let
\begin{equation}\label{hic7}
\psi_0=\psi_0(x,k(\lambda)) = e^{ik(\lambda)x}=e^{i (\sqrt{E}/2)  (\lambda \overline{z} + z/ \lambda)}, \quad \lambda \in (\mathbb C \backslash 0) \backslash T,
\end{equation}
where $k(\lambda)=k(\lambda,E)$ is defined by (\ref{kla}), $T$ is defined by (\ref{fir16}).

 We will denote by $G^+(\sqrt{E}), G(k)$ the convolution operators with kernels $G^+, G$ of (\ref{2DecAB}), (\ref{eq1}), and we will denote by $G^+(\sqrt{E})v, G(k)v$ the operators of multiplication by the potential $v$ followed by  convolution $G^+(\sqrt{E})$ or $G(k)$, respectively. Then, under the assumptions of lemma \ref{13NovA}, equations  (\ref{eq1}), (\ref{19JanA}) can be considered as linear integral equations for $\psi^+(\cdot,k)$,$\psi(\cdot,k) \in L^\infty(D)$, and can be rewritten as follows:
\begin{equation}\label{prp1}
\psi^+(\cdot,k)= (I-G^+(\sqrt{E})v)^{-1} \psi_0(\cdot,k), \quad
\psi(\cdot,k)= (I-G(k)v)^{-1} \psi_0(\cdot,k),
\end{equation}
for fixed $k \in \Sigma_E \backslash \Sigma_{E,\rho}$, where $I$ is the identity operator.

Thus
\begin{eqnarray}\label{hic8}
\psi^+(\cdot,k)= (I-G^+(\sqrt{E})v)^{-1} (I-G(k) v)\psi(\cdot,k), \\ \nonumber
\psi(\cdot,k) = (I- G^+(\sqrt{E})v)^{-1} (I- G^+(\sqrt{E})v) \psi(\cdot,k), \quad k \in \Sigma_E \backslash \Sigma_{E,\rho}.
\end{eqnarray}
Therefore,
\begin{equation}\label{prp}
\psi(\cdot,k)-\psi^+(\cdot,k) = (I-G^+(\sqrt{E})v)^{-1} (G(k) - G^+(\sqrt{E}))v\psi(\cdot,k) \quad k \in \Sigma_E \backslash \Sigma_{E,\rho}.
\end{equation}

We take $G - G^+$ from Lemma \ref{1DecLA} and use there that $\psi_0(x-y,k(\lambda))=\psi_0(x,k(\lambda))\psi_0(-y,k(\lambda))$. This leads to
\begin{eqnarray*}
(G(k(\lambda)) - G^+(\sqrt{E}))v \psi(\cdot,k(\lambda)) \\ = \frac{1}{(2\pi)^2}  \int_{D}\int_{\partial \Lambda} W(\lambda,\varsigma) \psi_0(x,k(\varsigma))\psi_0(-y,k(\varsigma))d\varsigma v(y) \psi(y,k(\lambda))dy\\=  \int_{\partial \Lambda} W(\lambda,\varsigma) \psi_0(x,k(\varsigma))h(k(\varsigma),k(\lambda))d\varsigma , \quad \lambda \in \partial \Lambda,
\end{eqnarray*}
where we used also (\ref{eq1303B}).

We plug the last relation in (\ref{prp}). It remains to note (see (\ref{prp1})) that \\ $(I-G^+(\sqrt{E}v))^{-1}\psi_0(\cdot,k(\varsigma)) =\psi^+(x,k(\varsigma)).$

\qed

\section{Proofs of Theorem \ref{thmir1} and Proposition \ref{prop41}}\label{sec6}
\subsection{Proof of Theorem \ref{thmir1}}
Let
\begin{eqnarray}\label{fr1}
\mu_0'(\lambda)=\mu'(\lambda)-e(\lambda), \quad \\ \nonumber
\mu_0^+(\lambda)=\mu^+(\lambda)-e(\lambda), \quad
\mu_0(\lambda)=\mu(\lambda)-e(\lambda),
\end{eqnarray}
where $\mu',\mu^+,\mu$ are the functions of (\ref{aprilA3}), $e(\cdot)$ is the function of (\ref{mir6}).

From formulas (\ref{mir3}), (\ref{mir6}) and from items 1 and 2 of proposition \ref{th13A} it follows, in particular,  that
\begin{eqnarray}\label{fr2}
\frac{\partial}{\partial \overline{\lambda}}e(\lambda)=r(x,\lambda,E) \overline{e(\lambda)}, \quad  \lambda \in \mathbb C ,\\ \label{fr3}
\frac{\partial}{\partial \overline{\lambda}}\mu'_0(\lambda)=r(x,\lambda,E) \overline{\mu'_0(\lambda)}, \quad \lambda \in \mathbb C \backslash \partial \Lambda, \\ \nonumber
\mu'_0(\lambda) \rightarrow 0 \mbox{ as }  \lambda \rightarrow \infty.
\end{eqnarray}
Proceeding from (\ref{fr3}) and using the generalized Cauchy formula for $\mu_0'$ (see formula (10.6) of chapter 3 of \cite{vekua})
one can obtain
\begin{equation}\label{fr4}
\mu_0'(\lambda)= \frac{1}{2\pi i } \int_{\partial \Lambda} \Omega_1(\lambda,\zeta) K_0(\zeta) d\zeta -\frac{1}{2\pi i } \int_{\partial \Lambda} \Omega_2(\lambda,\zeta) \overline{K_0(\zeta)} d\overline{\zeta}, \quad \lambda \in \mathbb C \backslash \partial \Lambda,
\end{equation}
where
\begin{equation}\label{fr6}
K_0(\lambda):= \mu_0^+(\lambda)- \mu_0(\lambda), \quad \lambda \in \partial \Lambda.
\end{equation}
In addition, from (\ref{mir12}), (\ref{fr1}) and (\ref{fr6})  it follows that
\begin{equation}\label{fr7}
K_0(\lambda)=K(\lambda), \quad \lambda \in \partial \Lambda.
\end{equation}
Formulas (\ref{fr1}), (\ref{fr4}), (\ref{fr7}) imply formula (\ref{mir11}).

Finally, equation (\ref{mir13}) follows from the substitution of (\ref{mir11}) into  (\ref{1DecC}) using formulas (\ref{mir4}), (\ref{aprilA3}), (\ref{mir12}), estimates (\ref{fr5}) - (\ref{zir2}) and the jump properties of the Cauchy integral.

This completes the scheme of proof of theorem \ref{thmir1}.

\subsection{Proof of Proposition \ref{prop41}}\label{sec72}
Equation (\ref{frl2}) follows from equation (\ref{mir13}) and formulas (\ref{frl1}), (\ref{frl1B}). Equation (\ref{fr12B}) follows from (\ref{frl2}).

Estimates (\ref{jun24A}), (\ref{Aug2A}) follow from formulas (\ref{mir3}), (\ref{mir4}), (\ref{frl1}), (\ref{frl1B}), estimates (\ref{zhir5}), (\ref{fr5}) - (\ref{zir3}), (\ref{zir7}), (\ref{zir8}) and the estimate
\begin{equation}\label{chg1}
\|\Omega^0_1 u \|_{L_p(\partial \Lambda)} \leq \const (p,\partial \Lambda) \|u\|_{L_p(\partial \Lambda)}, ~ 1<p<\infty,
\end{equation}
where
\begin{equation}\label{chg2}
(\Omega^0_1 u)(\lambda) = \frac{1}{2\pi i} \int_{\partial \Lambda} \frac{u(\varsigma)d\varsigma} {\varsigma - \lambda (1-0(|\lambda|-1))}, \quad \lambda \in \partial \Lambda,
\end{equation}
$u$ is a test function on $\partial \Lambda$. \qed

\section{Proofs of lemma \ref{fried1} and proposition \ref{nchg1}}\label{sec7}

\subsection{Proof of lemma \ref{fried1}}

Property (\ref{fried2}) follows from (\ref{chg2A}), (\ref{chg2C}).

Property (\ref{fried3}) follows from continuous dependence of $A_1,A_2$ with respect to $x\in \mathbb R^2$, $|x| \leq c$, at fixed $s \in ]\!-s_1,s_1[$ and
continuous dependence of $A_1,A_2$ with respect to $s \in ]\!-s_1,s_1[$ uniformly in $x \in \mathbb R^2$, $|x| \leq c$, in the sense of $\|\cdot\|_{L^2(\partial \Lambda \times \partial \Lambda)}$, for fixed $c>0$.

In turn, these continuities of $A_1,A_2$ in $x$ and in $s$ follow from formulas (\ref{zir1}), (\ref{zir2}), (\ref{frl1B}) and the following results:

 (i) $h|_{\partial \Lambda \times  \partial \Lambda}$  depends continuously on $s \in ]-s_1,s_1[$  in the  sense of $\|\cdot \|_{C(\partial \Lambda \times  \partial \Lambda)}$,

(ii) $u_{E,\rho}$ depends continuously on $s \in ]-s_1,s_1[$ in  the  sense of $\|\cdot\|_{L_{p,2}(\mathbb C)}, ~ 2<p<4,$
where $h=h(k(\lambda),k(\varsigma)), ~ u_{E,\rho}$ correspond  to $sv$ according to  (\ref{19JanA})-(\ref{air2}), (\ref{eq1303B}), (\ref{kla}), (\ref{fir20}), (\ref{mir14});

(iii) The following estimates hold:
$$
\left | e^{-2i\Re k(\lambda)x} - e^{-2i\Re k(\lambda)x'} \right | \leq \Const \cdot
(\sqrt{E} (|\lambda|+ |\lambda|^{-1}  )|x-x'|)^{\alpha}, ~ \lambda  \in \mathbb C \backslash 0, ~ x,x' \in \mathbb R^2, ~ 0<\alpha\leq 1,
$$
$$
\left | e^{i(k(\varsigma)-k(\lambda))x} -  e^{i(k(\varsigma)-k(\lambda))x'} \right | \leq 2(E+2\rho^2)^{1/2} e^{2\rho \max(|x|,|x'|)}|x-x'|, ~ \varsigma,\lambda \in \partial \Lambda, ~ x,x' \in \mathbb R^2;
$$

(iv)  If $u \in L_{p,2}(\mathbb C), 2<p<4$, then $(|\lambda|+|\lambda^{-1}|)^{\alpha} u(\lambda) \in L_{p',2}(\mathbb C)$ (as a function of $\lambda$), $2<p'<p(1+\alpha p /2)^{-1}$, where $0<\alpha<(p-2)/p;$

(v) The map (defined via (\ref{mir6}))
$$
r \in L_{p,2}(\mathbb C) \rightarrow e(\cdot) \in C(\mathbb C)
$$
is continuous and the maps (defined via (\ref{mir7}), (\ref{mir8}))
$$
r \in L_{p,2}(\mathbb C) \rightarrow X_j \in C(\mathbb C^2 \backslash \mathbb C_\varepsilon), ~ j=1,2,
$$
$$
\mathbb C_\varepsilon = \{ (\lambda,\varsigma) \in \mathbb C^2  ~ : ~ |\lambda-\varsigma|<\varepsilon \},
$$
are continuous for any $\varepsilon >0$, where $L_{p,2}(\mathbb C)$ is considered with the norm of (\ref{mir15}), $2<p<4$, and
$C(\mathbb C),C(\mathbb C^2 \backslash \mathbb C_\varepsilon)$ are considered with the uniform  norms.

In order to prove (\ref{fried4}) we consider $sv$, where $s \in \mathbb C$,  and we consider $h_s = h_s(k(\lambda),k(\varsigma)),$  $\lambda, \varsigma \in \partial \Lambda $, and $b_s =b_s(k(\lambda)), \lambda \in (\mathbb C \backslash 0 )\backslash \Lambda$, where $h_s,b_s$ correspond to $sv$ according to (\ref{19JanA})-(\ref{air2}), (\ref{eq1303B}), (\ref{kla}), (\ref{fir20}) (with $sv$ in place of $v$).  Proceeding from these formulas and equations and from (\ref{air3}), (\ref{fir15}), (\ref{fir24}),  one can show that there is
an open neighbourhood $\mathcal N$ of the real interval $]-s_1,s_1[$ in $\mathbb C$ (where $\mathcal N$ depends on $D,\|v\|_{L^\infty(D)},E,\rho,q$) such that \begin{equation}\label{chg19A}
\overline{\mathcal  N} =\mathcal N, \mbox{ i.e. } \mathcal N \mbox{ is symmetric with respect to } \mathbb R,
\end{equation}
\begin{eqnarray}\label{chg19B}
h_s(\cdot,\cdot,E) \in C(\partial \Lambda \times  \partial \Lambda), ~ u_{E,\rho,s} \in L_{p,2} (\mathbb C), ~ 2<p<4,  \\ \nonumber
\mbox{with holomorphic dependence on } s \in \mathcal N,
\end{eqnarray}
where $u_{E,\rho,s}$ is defined by (\ref{mir14}) with $b_s$ in place of $b$.

Next, we consider $e_s, X_{1,s}, X_{2,s}, \Omega_{1,s}, \Omega_{2,s}$ defined via (\ref{mir6}), (\ref{mir7}), (\ref{mir8}), (\ref{mir9}) with $r_s$ in place of $r$, where $r_s$ is defined by (\ref{mir3}) with $b_s$ in place of $b$, where $s \in ]-s_1,s_1[$. And we consider $e^{\pm}_{s,\sigma}$, $X^{\pm}_{j,s,\sigma},~j=1,2$, defined via the following systems of equations:
\begin{eqnarray}\label{chgmir6}
e^+_{s,\sigma}(\lambda)=1- \frac{1}{\pi} \int_{\mathbb C} r_s(x,\zeta,E) {e^-_{s,\sigma}(\zeta)} \frac{d\Re \zeta d\Im \zeta}{\zeta- \lambda}, \\ \nonumber
e^-_{s,\sigma}(\lambda)=1- \frac{1}{\pi} \int_{\mathbb C} \overline{r_{\overline \sigma}(x,\zeta,E)} {e^+_{s,\sigma}(\zeta)} \frac{d\Re \zeta d\Im \zeta}{\overline \zeta- \overline \lambda},
\end{eqnarray}
\begin{eqnarray}\label{chgmir7}
X^+_{1,s,\sigma}(\lambda,\zeta) + \frac{1}{\pi} \int_{\mathbb C} r_s(x,\eta,E) {X^-_{1,s,\sigma}(\eta,\zeta)}\frac{d\Re \zeta d\Im \zeta} {\eta - \lambda}= \frac{1}{2(\zeta- \lambda)}, \\ \nonumber
X^-_{1,s,\sigma}(\lambda,\zeta) + \frac{1}{\pi} \int_{\mathbb C} \overline{r_{\overline \sigma}(x,\eta,E)} {X^+_{1,s,\sigma}(\eta,\zeta)}\frac{d\Re \zeta d\Im \zeta} {\overline \eta - \overline \lambda}= \frac{1}{2(\overline \zeta- \overline \lambda)},
\end{eqnarray}
\begin{eqnarray}\label{chgmir8}
X^+_{2,s,\sigma}(\lambda,\zeta) + \frac{1}{\pi} \int_{\mathbb C} r_s(x,\eta,E) {X^-_{2,s,\sigma}(\eta,\zeta)}\frac{d\Re \zeta d\Im \zeta} {\eta - \lambda}= \frac{1}{2i(\zeta- \lambda)}, \\ \nonumber
X^-_{2,s,\sigma}(\lambda,\zeta) + \frac{1}{\pi} \int_{\mathbb C} \overline{r_{\overline \sigma}(x,\eta,E)} {X^+_{2,s,\sigma}(\eta,\zeta)}\frac{d\Re \zeta d\Im \zeta} {\overline \eta - \overline \lambda}= \frac{-1}{2i(\overline \zeta- \overline \lambda)},
\end{eqnarray}
where $s,\sigma  \in \mathcal N$, $~r_s$ is defined by (\ref{mir3}) with $b_s$ in place of $b$.  In addition, we consider also
\begin{equation}\label{Setmir9}
\Omega_{1,s,\sigma}(\lambda,\zeta) := X^+_{1,s,\sigma}(\lambda,\zeta) + i X^+_{2,s,\sigma}(\lambda,\zeta), \quad \Omega_{2,s,\sigma}(\lambda,\zeta) := X^+_{1,s,\sigma}(\lambda,\zeta) - i X^+_{2,s,\sigma}(\lambda,\zeta),
\end{equation}
where $\lambda,\zeta \in \mathbb C,$ $s,\sigma \in \mathcal N.$

Let
\begin{equation}\label{Setmir9B}
S := \left\{ (s,\sigma)\in \mathcal N \times \mathcal N  ~ : ~ \sigma=s \in ]-s_1,s_1[ ~ \right\}.
\end{equation}

Using considerations of section 9 of chapter 3 of \cite{vekua}, one can show that systems (\ref{chgmir6}), (\ref{chgmir7}), (\ref{chgmir8}) for $e^{\pm}_{s,\sigma},X^{\pm}_{j,s,\sigma},j=1,2$, for $(s,\sigma)\in S$, are reduced to the equations for $e_s$, $X_{j,s}, j=1,2$, $s\in]-s_1,s_1[,$ are uniquely solvable in $L^q_0(\mathbb C), p/(p-1) \leq q <2, $ where $p$ is the number (\ref{chg19B}).  In addition:
\begin{equation}\label{Setmir10}
e_s=e^+_{s,s}, \quad \overline{e_s}= e^-_{s,s},  \quad X_{j,s}=X^+_{j,s,s}, \quad \overline{X}_{j,s}=X^-_{j,s,s}, \quad \Omega_{j,s}=\Omega_{j,s,s},
\end{equation}
where $j=1,2,$ $s\in ]-s_1,s_1[.$

Using the definition of $r_s$ and holomorphic dependence of $u_{E,\rho,s}$ on $s \in \mathcal N$ in (\ref{chg19B}) one can show that
\begin{eqnarray}\label{Setmir11}
r_s (x,\cdot,E) \in L_{p,2} (\mathbb C), \quad \overline{r_{\overline{\sigma}}(x,\cdot,E)} \in L_{p,2}(\mathbb C), \quad 2<p<4, \\ \nonumber
\mbox{with holomorphic dependence on } s,\sigma \in \mathcal N,
\end{eqnarray}
for fixed $x\in \mathbb R^2, E>0$.

Proceeding from these results and from properties of the integral operators in (\ref{chgmir6}) -(\ref{chgmir8})  (presented in \cite{vekua}), one can show that there is an open neighbourhood $\mathcal S_x$ of $S$ in $\mathcal N \times \mathcal N$ (where $\mathcal S_x$ depends also on $v,E,\rho$)  such that:
\begin{eqnarray}\label{Setmir12}
\mbox{systems (\ref{chgmir6}), (\ref{chgmir7}), (\ref{chgmir8}) for } e^{\pm}_{s,\sigma}, X^{\pm}_{j,s,\sigma}, j=1,2, \mbox{are uniqely solvable in } L_{q,0} (\mathbb C), \\ \nonumber
p/(p-1) \leq q <2, \mbox{ for } (s,\sigma) \in \mathcal S_x;
\end{eqnarray}
\begin{eqnarray}\label{Setmir13}
e^+_{s,\sigma} \in C(\mathbb C), ~ \Omega_{j,s,\sigma} \in C(\mathbb C^2 \backslash \mathbb C_\varepsilon), ~j=1,2,  \mbox{ for any } \varepsilon >0,\\ \nonumber
\mbox{with holomorphic dependence on } (s,\sigma) \in \mathcal S_x,
\end{eqnarray}
where $\mathbb C_\varepsilon$ is defined in item (v) in the proof of property (\ref{fried3});
\begin{eqnarray}\label{Setmir14}
\left | \Omega_{1,s,\sigma}(\lambda,\zeta) - \frac{1}{\zeta - \lambda} \right | <  \frac{c_3(s,\sigma,p)}{|\zeta - \lambda|^{2/p}},
\quad
\left | \Omega_{2,s,\sigma}(\lambda,\zeta) \right | < \frac{c_3(s,\sigma,p) }{|\zeta - \lambda|^{2/p}},
\end{eqnarray}
where $c_3$ depends continuously on $(s,\sigma) \in \mathcal S_x$  and depends also on $v$.

Let
\begin{eqnarray}\label{Setmir15}
\mathcal N_x := \{ s \in \mathcal N ~ : ~ (s,s) \in \mathcal S_x ~ \}, ~ x \in \mathbb R^2.
\end{eqnarray}
One can see that $\mathcal N_x$ is an open neighbourhood of the real interval $]-s_1,s_1[$ in $\mathbb C$.

We consider
\begin{eqnarray}\label{Setmir16}
A_{1,s}(\lambda,\zeta)= A_{1,s}(x,\lambda,\zeta,E) =  \frac{1}{2\pi i } \int_{\partial \Lambda} \!R_s(x,\lambda,\lambda',E)  \Omega_{1,s,s}(\lambda'(1\!-0(|\lambda'|-1)),\zeta)
d\lambda ', \\ \nonumber
A_{2,s}(\lambda,\zeta)= A_{2,s}(x,\lambda,\zeta,E) =  \frac{-1}{2\pi i } \int_{\partial \Lambda} R_s(x,\lambda,\lambda',E)  \Omega_{2,s,s}(\lambda',\zeta)
d\lambda ', \quad \lambda,\zeta \in \partial \Lambda,
\end{eqnarray}
where $R_s$ is defined by (\ref{mir4}) with $h_s$ in place of $h$, ~$\Omega_{1,s,\sigma},\Omega_{2,s,\sigma}$ are the functions of (\ref{Setmir9}), (\ref{Setmir13}), (\ref{Setmir14}),
$\lambda,\zeta \in \partial \Lambda,$ $s \in \mathcal N_x$.

We consider also
\begin{eqnarray}\label{Setmir17}
\widetilde{A}_{j,s} := \overline{A_{j,s}}, ~ j=1,2, ~ s \in \overline{\mathcal N_x}.
\end{eqnarray}
Using (\ref{chg19B}) for $h_s$ and (\ref{Setmir13}), (\ref{Setmir14}) for $\Omega_{j,s,s},j=1,2$, we obtain
\begin{eqnarray}\label{Setmir18}
A_{j,s} \in L_2(\partial \Lambda \times \partial \Lambda), ~ j=1,2, ~ \\ \nonumber
\mbox{ with holomorphic dependence on } s\in \mathcal N_x.
\end{eqnarray}
Using (\ref{Setmir17}), (\ref{Setmir18}) we also obtain
\begin{eqnarray}\label{Setmir19}
\widetilde{A}_{j,s} \in L_2(\partial \Lambda \times \partial \Lambda),~ j=1,2, \\ \nonumber
\mbox{ with holomorphic dependence on } s\in \overline{\mathcal N_x}.
\end{eqnarray}

We  consider $A(x,s)$, where $s \in \mathcal N_x \cap \overline{\mathcal N_x}$, defined using (8.14), (8.15) in a similar way with $A(x,s)$ for $s \in ]-s_1,s_1[$, but with $\widetilde{A}_{j,s}$ in place of $A(x,s)$. Finally, we consider $\det A(x,s)$ for $s \in \mathcal N_x \cap \overline{\mathcal N_x}$.

Using  (\ref{Setmir10}) for $\Omega_{j,s,s},$ (\ref{Setmir18}), (\ref{Setmir19}), we obtain that
\begin{eqnarray}\label{Setmir20}
\det A(x,s) \mbox{ is holomorphic in } s \in \mathcal N_x \cap \overline{\mathcal N_x} \mbox{ for fixed } x \in \mathbb R^2.
\end{eqnarray}
Property (\ref{Setmir20}) implies property (\ref{fried4}). \qed

\subsection{Proof of proposition \ref{nchg1}}\label{sec82}
Let $v$ be as in remark \ref{nchg2} and let $\det A(x,s)$ be defined like in lemma \ref{fried1}.

Let
\begin{eqnarray}\label{Setmir21}
Z := \{ (x,s) \in \mathbb R^2 \times ]-s_1,s_1[ ~ :~ \det A(x,s) =0 ~ \}, \\ \nonumber
Z_x := \{ s \in ]-s_1,s_1[ ~: ~ \det A(x,s) =0 ~ \}, ~ x \in \mathbb R^2, \\ \nonumber
Z_s := \{ x  \in \mathbb R^2 ~: ~ \det A(x,s) =0 ~ \}, ~ s \in  ]-s_1,s_1[. \\ \nonumber
\end{eqnarray}
Using (\ref{fried2}), (\ref{fried4}), we obtain that $Z_x$ is a discrete set (maybe empty) without interior accumulation points in interval $]-s_1,s_1[$.
Therefore, we have, in particular, that
\begin{eqnarray}\label{Setmir22}
\rm{Meas } ~Z =0 \mbox{ in }\mathbb R^2\times ]-s_1,s_1[.
\end{eqnarray}
As a corollary,
\begin{eqnarray}\label{Setmir23}
\rm{Meas } ~Z_s =0 \mbox{ in }\mathbb R^2 \mbox{ for almost each } s \in ]-s_1,s_1[.
\end{eqnarray}
Property  (\ref{Setmir23}) implies the result of proposition \ref{nchg1} interpreted according to remark \ref{nchg2}. \qed

{\bf Acknowledgments.} E.L. is grateful to  U. Kahler, A.Slunyaev, A. Gelash, S.Turitsyn for useful discussions.

\end{document}